\documentclass[debug,overfull]{epl}
\usepackage{epsfig}
\def\ltsim{\vbox {\hbox{\lower .8\baselineskip \hbox{$<$}} \break
                 \hbox{\lower 0.2\baselineskip \hbox{$\sim$}} } }

\title{Interference Blockade in the Conductance \\ of Organic  Molecules}

\shorttitle{Interference ...}
\author{R. Gutierrez, F. Grossmann and R. Schmidt}
\institute{
Institute for Theoretical Physics,
Technical University of Dresden, D-01062 Dresden, Germany\\
}


\begin{document}

\maketitle
\begin{abstract}
The conductance of {\em cis/trans} isomers of stilbene molecules
connected to armchair single wall carbon nanotubes
is studied in
the Landauer formalism combined with a density-functional based
approach. For a given arrangement of the electrodes,
dramatic differences in the transmission between both isomers are found.
For a given
isomer, the conductance can be varied by orders of magnitude by
changing the molecule-electrode relative orientation. Both effects can be
explained by a simple, physically transparent interference rule, which suggests 
a straightforward conductance control
in such molecular systems by different switching mechanisms.

\end{abstract}

{\it Introduction}. - In recent years rapid progress in the field of molecular electronics has been 
made. Novel experimental approaches allow the investigation of the electronic 
transport properties of small molecular groups or even single molecules  connected 
to macroscopic or mesoscopic electrodes
\cite{Review}. Effects like negative differential resistance (NDR)
\cite{chen99,ra02,hett02} and rectification \cite{KHY02,LA02,TBS02}
have been demonstrated. 
Common to this class of systems is the  subtle interplay between electronic and 
structural properties in determining the electronic transport. This makes 
the search for molecules that exhibit
{\it controllable} conformational changes highly desirable. 
Thus, recent experiments on conjugated organic molecules have shown 
reversible conductance 
switching which can be related to reorientation of single molecules induced by 
voltage pulses \cite{gao00}, to internal structural modifications 
induced by charge transfer within the molecular complex \cite{ma98,Retal01} 
or to conformational changes due to interactions with 
the local environment \cite{don01}. 

Stilbene ((1,2)-diphenyl-ethylene) is a  prototypical example of a 
system with a controllable transition between two 
stable states. This molecule 
undergoes a {\em cis-trans} isomerization  around the central 
ethylenic bond under, e.g.  the influence of a laser field \cite{stil1}, which 
suggests a natural way to realize a switching mechanism. 
Many experimental and theoretical investigations have been carried out in order to
understand the electronic and structural properties of the molecule, 
the mechanism leading
to isomerization as well as its optimal control \cite{stil1,stil2,stil3,gross02}.
Less theoretical attention has been paid, however, 
to the electronic transport properties of 
the isomers and, especially, to the possibility of identifying them via 
their fingerprints in the conductance spectrum. 

In this Letter we investigate electron transport in {\em cis/trans} stilbene 
within a pure carbon molecular device, in which the molecule 
is covalently bonded to two carbon nanotubes (CNT) 
acting as electrodes. The possibility to develop a carbon based 
nanoelectronics has been investigated theoretically \cite{gutie02} 
and in recent transport experiments \cite{Retal01}. 
We will show that for some specific molecule-electrode orientations
the conductance of both isomers is drastically suppressed around the 
Fermi energy and that this effect can be clearly related to quantum interference 
of the 
electronic waves ("interference
blockade").  The appearance or 
disappearance of the blockade effect can be simply controlled by the Fermi 
wave vector $k_F$ of the electrodes 
and the difference in the geometrical pathways $\Delta x$ of 
the electronic wave functions around the phenyl rings. 
The following simple relation holds for the studied molecular 
junctions: 
\begin{eqnarray}
k_F \Delta x\approx n\pi . 
\end{eqnarray}
Blockade occurs for odd n ("off-state") while nearly ideal transmission 
is found for even n ("on-state"). Switching between both 
conductance states can be controlled 
either by a conformational change, 
e.g.~by {\em cis/trans} isomerization 
or by reorientation of the whole molecule with respect to the electrodes, 
in accord with recent experimental observations on similar systems \cite{gao00,Retal01,don01}.

\ \\

{\it Theoretical Method}. - Our computational approach \cite{pra01}  
combines a density-functional-based
tight-binding (DF-TB) formalism with numerical Green function techniques
to investigate electronic transport within the Landauer theory.
The basic quantity to be calculated in the following is the
two-terminal conductance $g=\frac{e^2}{\pi\hbar} T(E_F)$, 
which is proportional to the
transmission probability $ T(E_F)$ at the equilibrium Fermi energy $E_F$
in the  linear response regime and at zero temperature \cite{Datt95}.  
The transmission can be expressed as:
\begin{eqnarray}
T(E)=4{\rm Tr}[ \Im m{\bf\Sigma}_{L}\, {\bf G} \,\Im m{\bf\Sigma}_{R}\, {\bf G}^{\dagger}].
\end{eqnarray}
Here, ${\bf G}$ is a retarded molecular Green function
including  self-energy interactions ${\bf \Sigma}_{L,R}$ with the left (L) and 
right (R) electrodes.
The self energies  are given by 
${\bf \Sigma}_{L,R}={\bf V}_{L,R}^\dagger {\bf g}_{L,R} {\bf V}_{L,R}$, where  
${\bf V}_{L,R}$ are electrode-molecule coupling terms and 
${\bf g}_{L,R}$ is the retarded Green function of the electrodes, which is 
obtained using recursive techniques \cite{LLR85}.
In calculating these quantities in the DF-TB scheme
we use a nonorthogonal basis set 
including the 2s2p carbon valence orbitals and the 1s orbitals of hydrogen. 

Upon coupling to the electrodes the molecular orbitals are broadened
and partially shifted, so that instead of looking at the electronic eigenvalue
spectrum of the isolated molecules, it is more useful to analyze
the projected density of
states (PDOS), obtained from the total DOS of the device by a partial trace:
\begin{eqnarray}
\nu_l(E)=-\frac{1}{\pi N_l} \Im m \, Tr_l [{\bf G}(E){\bf S}].
\end{eqnarray}
In this expression the index $l$ runs over some appropriate subset containing $N_l$ 
atomic sites and the overlap matrix ${\bf S}$ takes into account the non-orthogonality of the basis set.
\ \\

{\it Electronic Transport}. - 
The molecular devices are shown in the upper panel of Fig.~1. 
Two armchair (5,5) CNTs \cite{apl} with open ends are bridged by a 
single stilbene isomer. 
In the free molecules 
the {\em trans} state is almost planar, while in the {\em cis} configuration
the phenyl
groups undergo a combination of twisted and torsional
rotations around the C=C double
bond, leading to a more compact structure.
As shown in Fig.~1 the CNTs are oriented in a way that their axes are parallel 
to each other. 

In order to improve the molecule-electrode 
contact anchoring groups are 
usually used, e.g. thiol groups on gold surfaces \cite{datta97}. 
Here, we want to study  a carbon based setup, so that CH radicals are used 
as anchoring groups\cite{comment0}. In addition, we limit our discussion 
to the case of a single contact point 
between the molecule and the CNT surfaces 
(if not stated otherwise, when speaking of {\em trans} and {\em cis} 
stilbene we refer always to the (CH)$_2$-stilbene system). 
To simulate the device's atomic structure we use a cluster 
consisting of the molecule and eight unit cells of the CNT on either side of it.
Structural optimization is performed by using  conjugate-gradient
techniques. Hereby, only the molecule and the surface unit cell of the
tubes are allowed to relax.
The bond length of the CH-CNT 
single contact was determined to lie around 1.42 \AA~ depending on the device 
geometry, a value intermediate between single and double bond, which 
guarantees a strong covalent bond to the electrodes. 

\vspace{1.8cm}
\begin{minipage} [r] {8.5cm}
\hspace{+2.9cm}{
\raisebox {+4.55cm}[+5.55cm]
{\epsfig{figure=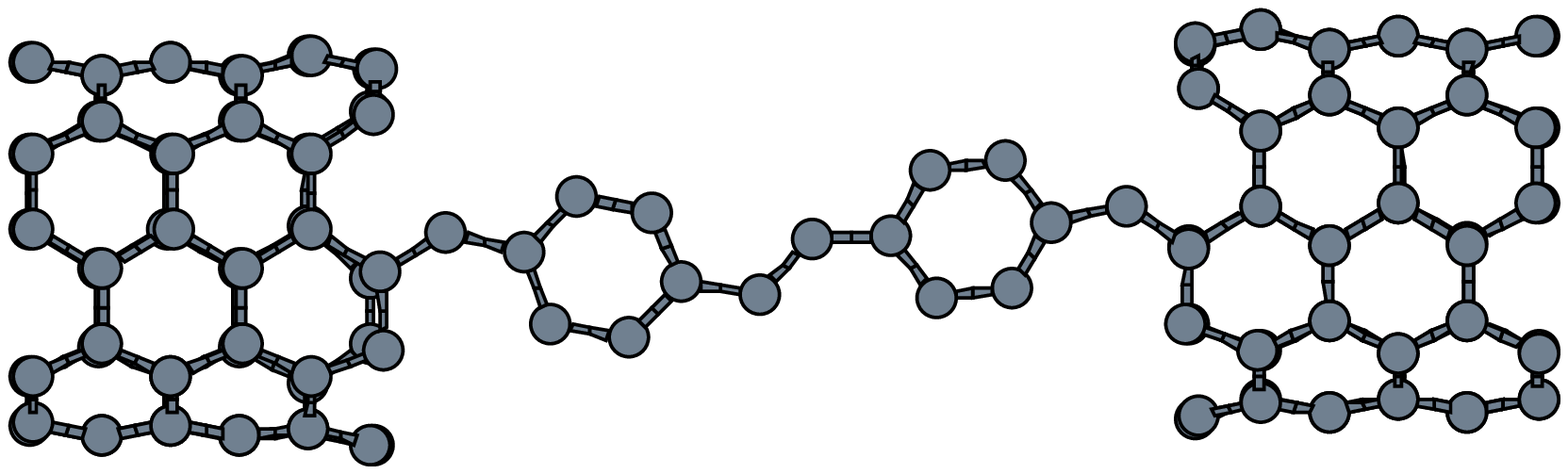, width=4.0cm, height=3.0cm}}
}
\end{minipage}
\hfill
\begin{minipage} [c] {4.9cm}
\hspace{-2.0cm}{
\raisebox {+4.95cm}[+5.75cm]
{\epsfig{figure=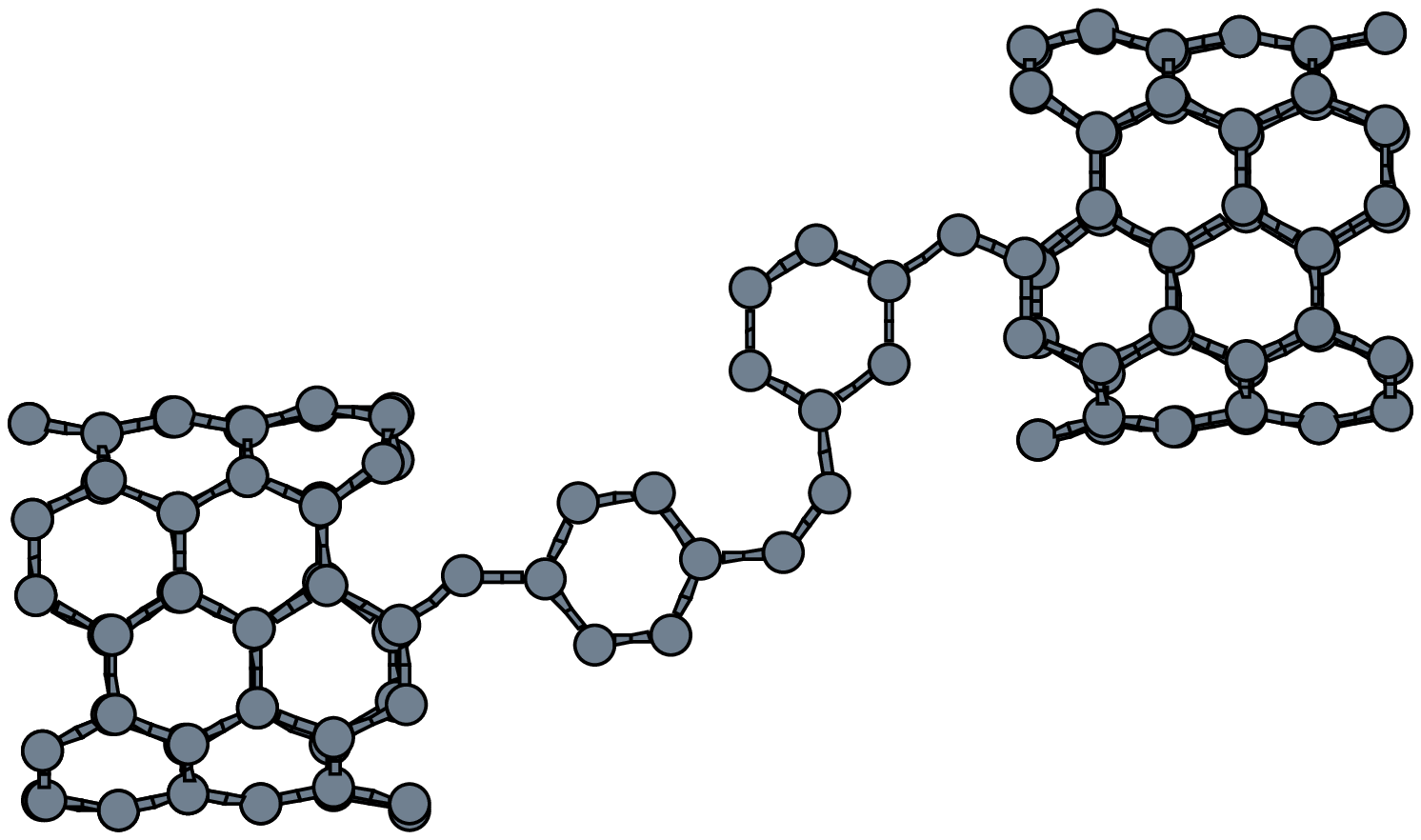, width=4.0cm, height=3.0cm}}
}
\end{minipage}

{
\vspace{-6.2cm} {
\begin{figure} [h]
\begin{center}
\hspace{-0.4cm}
\epsfig{figure=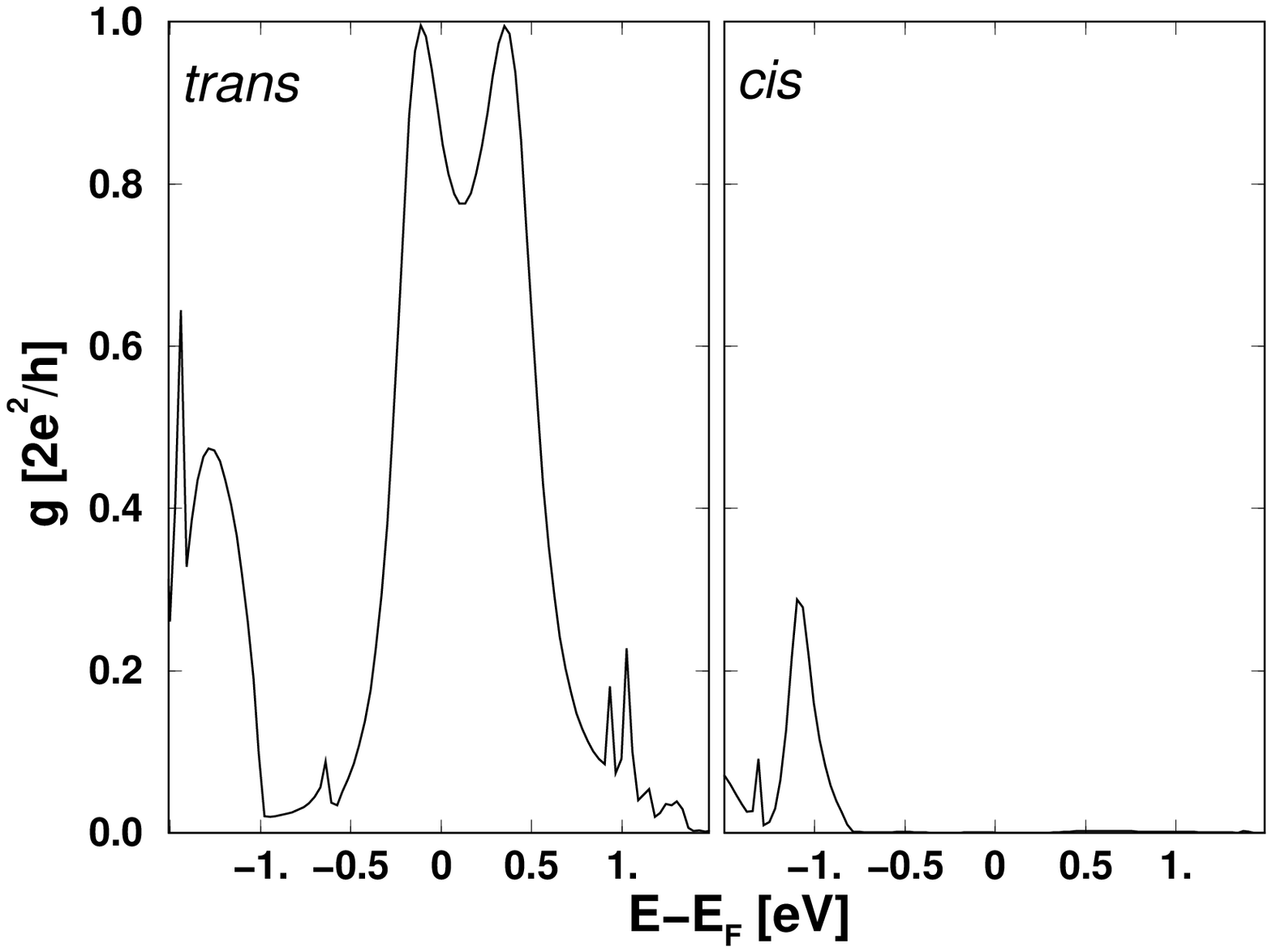, width=8.5cm, height=5.8cm}
\caption{ Conductance of {\em trans} and {\em cis} stilbene for the same
electrode arrangement. Only the carbon frame is shown.
The conductance is strongly suppressed around the Fermi energy for the  {\em cis}
isomer.
}
\end{center}
\end{figure}
}}

In the lower panel of Fig.~1 the conductance $g$ in units of $2e^2/h$ is plotted as a 
function of the incoming electron energy. 
We found a drastic difference in the transmission around the Fermi energy. While for 
{\em trans}-stilbene two large resonances are seen with almost 
perfect transmission $T\approx 1$ the {\em cis} state
shows almost no transmission in the same spectral window. 
This is quite surprising as both isomers posses electronic states near 
the Fermi level, as will be shown in the following.  Moreover, 
it should be noted, that the large level broadening due to strong 
molecule-electrode coupling precludes 
Coulomb blockade effects.

In Fig.~2 we show the PDOS $\nu_l$, as defined in Eq.~3. The index $l$ runs 
over molecular sites and over CNT surface sites, separately. 
The vertical arrows indicate the position of the molecular orbitals 
of the  isomers previous to contacting them to the CNTs. 
The indicated lowest unoccupied (LU) molecular orbital (MO) LUMO and the 
LUMO+1 are 
$\pi$-orbitals with a large contribution from 
the C-atoms belonging to the anchoring groups. 
The LUMO and LUMO+1 of the {\em cis}  isomer
are almost degenerate. The amount of splitting depends however on the position 
of the CH-radicals along the phenyl ring.  
The large resonances near the Fermi energy in both isomers are clearly related to 
molecular orbitals, because the surface CNT-PDOS 
is rather featureless in this energy region. Side peaks around -0.5 eV and between 
1--1.5 eV 
are mainly determined by CNT states which derive from surface dangling bonds. 
These states are  
originally much closer to the Fermi level, but upon relaxation they move away from 
the midgap due to the slight restructuration of the CNT surface. Moreover, 
the highest occupied MO (HOMO) cannot 
be clearly identified because it lies in an energy window where bulk states
are also present, so that both states can strongly mix.

\begin{figure}
\begin{center}
\epsfig{figure=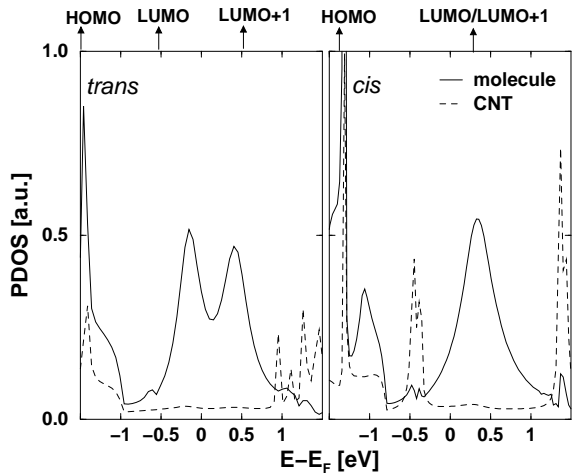, width=7.5cm,height=6.5cm }
\caption{
Density of states (in arbitrary units),
projected onto the
molecule (solid line)  and onto the CNT surface atoms (dashed line) for
the {\em trans} and {\em cis} configurations shown in Fig.~1, upper panel.
The HOMO, LUMO and LUMO+1 of the
isolated molecules ({\em including} the CH-radicals)
are shown as a guide for the eye.
Due to the strong electrode-molecule coupling the molecular resonances are shifted
 and broadened.
  }
\end{center}
\end{figure}

The main point to be emphasized here is, however, that {\em cis} stilbene 
shows a prominent 
resonance (the nearly degenerated LUMO and LUMO+1 states) near the 
Fermi energy, which 
nevertheless does not show up in the transmission. 
Thus, the strong conductance suppression
cannot be related to an intrinsic molecular property and, hence it must be related 
to the interfacial topology. 

To analyze this in more detail, we have considered several possible {\em single} 
contact geometries of 
{\em trans}--stilbene which are schematically shown in Fig.~3. Note that we keep 
the left electrode fixed and  change only the contact points on the right side. 
The transmission at the Fermi energy (presented in the lower panel of  Fig.~3) 
as a function 
of the contact point shows an oscillating behavior with strong conductance 
suppression by about 3 orders of magnitude at positions (b) and (d). 

A qualitative understanding of this interference blockade follows from the 
different paths around the 
right phenyl ring, an electron can take. 
For position 
(c) e.g., there is no difference between the clockwise and counterclockwise paths, 
whereas for position (b) a path difference of about 
$\Delta x\approx 3$ {\rm \AA}
(two side lengths of the hexagon) does exist. This  leads to a phase difference
of approximately $k_F\Delta x\approx3\approx\pi$ (where we have used
the calculated Fermi wave vector of $k_F\approx 1{\rm \AA}^{-1}$), 
which makes the paths interfere almost
completely destructively. Thus, by moving around the phenyl 
ring we are basically 
inducing a change in the phase of the electronic waves, leading finally to  
Eq. (1). It should be noted that the dramatic but periodic conductance variation 
is clearly related to the existence 
of only two possible paths for the electrons along the molecular frame. 
It is remarkable that the qualitative explanation given above is quite insensitive 
to structural relaxation. 
The absolute values of the transmission are of course influenced 
by the latter, but not the general trend shown in Fig.~3.


\begin{figure}[h]
\begin{center}
\hspace{1cm}\epsfig{figure=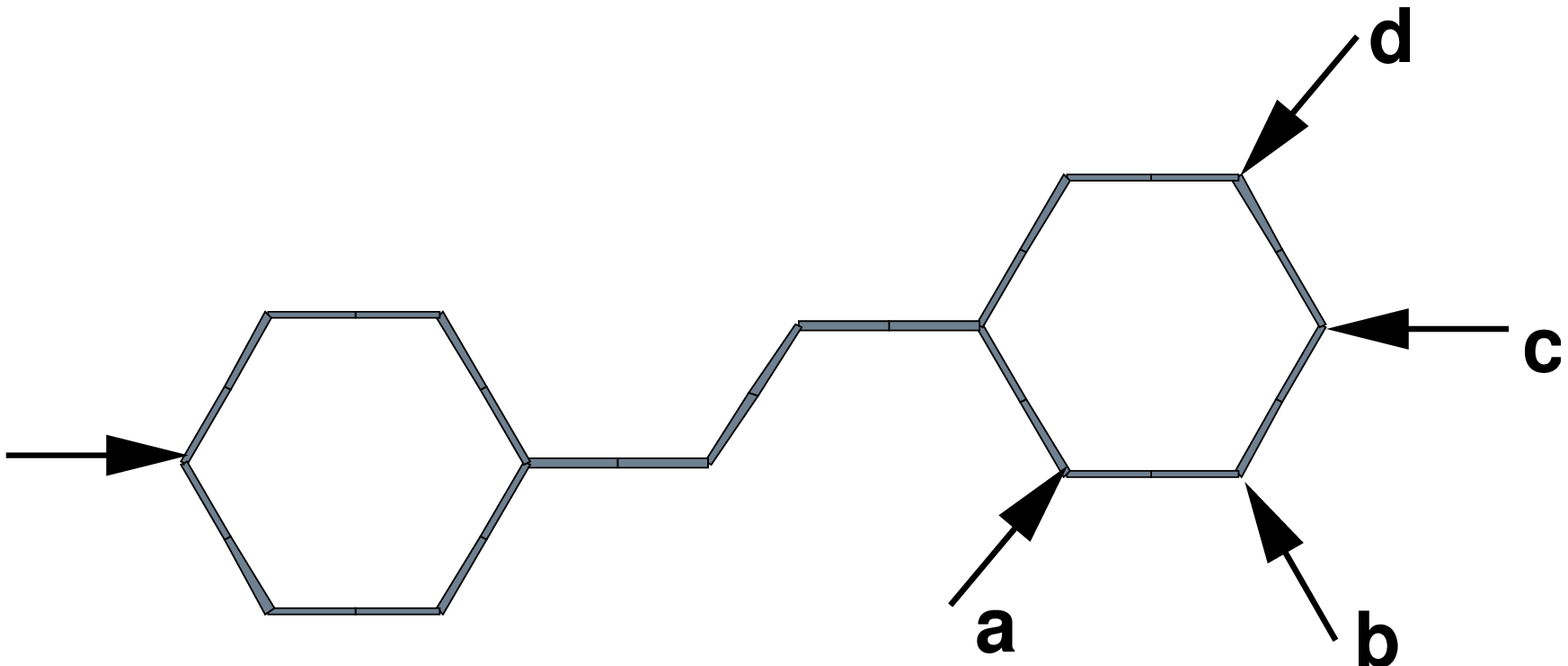, width=4.0cm } \\
\epsfig{figure=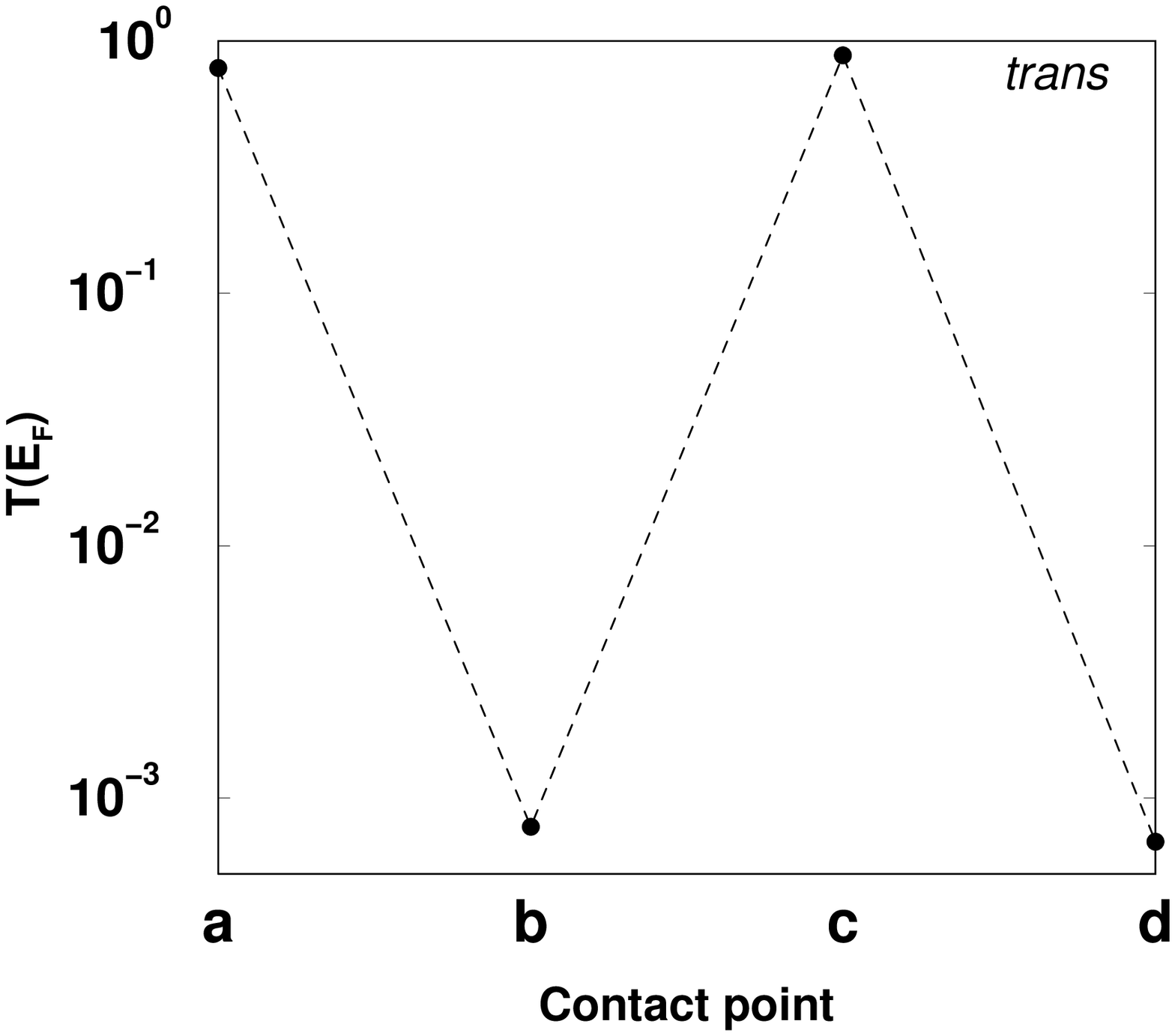, width=5.9cm }
\end{center}
\caption{ Transmission at the Fermi energy for {\em trans} isomer for
different (single) contact points (schematically shown in the upper panel)
of the right CNT-electrode to the molecule. Arrows indicate the position of the
CH-radical. A change of the contact point
modifies the relative phase of electrons moving around the phenyl rings.
}
\end{figure}

If the above arguments have some generality they should  
hold also for {\em cis} stilbene. 
By modifying the contact points (i.e.~the 
position of the CH-group on one of the phenyl ring according to the scheme 
of Fig.~3), 
it should therefore  be possible to "switch on" the molecular 
resonances of the {\em cis} isomer. This is 
indeed the case, see Fig.~4. The contact points for the {\em trans} and 
{\em cis} isomers correspond to the cases (d) and (c) of Fig.~3, respectively. 
While the transmission in the {\em trans} system is suppressed, the 
LUMO/LUMO+1 resonances (not well resolved due to strong coupling)  
of the {\em cis} isomer are now clearly visible in the conductance.

Finally, we note that the influence of internal molecular 
symmetries as well as contact effects, on the 
electronic transport, have been recently addressed 
experimentally\cite{KHY02,reich02}
and theoretically \cite{ra02,hett02,TBS02,fagas01}.
The molecular junction studied here combines both aspects, as the CH-radicals 
not only modify the symmetry of the stilbene molecule but, at the same time, 
they determine the contact point to the electrodes.

\begin{minipage} [l] {9.9cm}
\hspace{+3.5cm}{
\raisebox {+3.75cm}[+5.55cm]
{\epsfig{figure=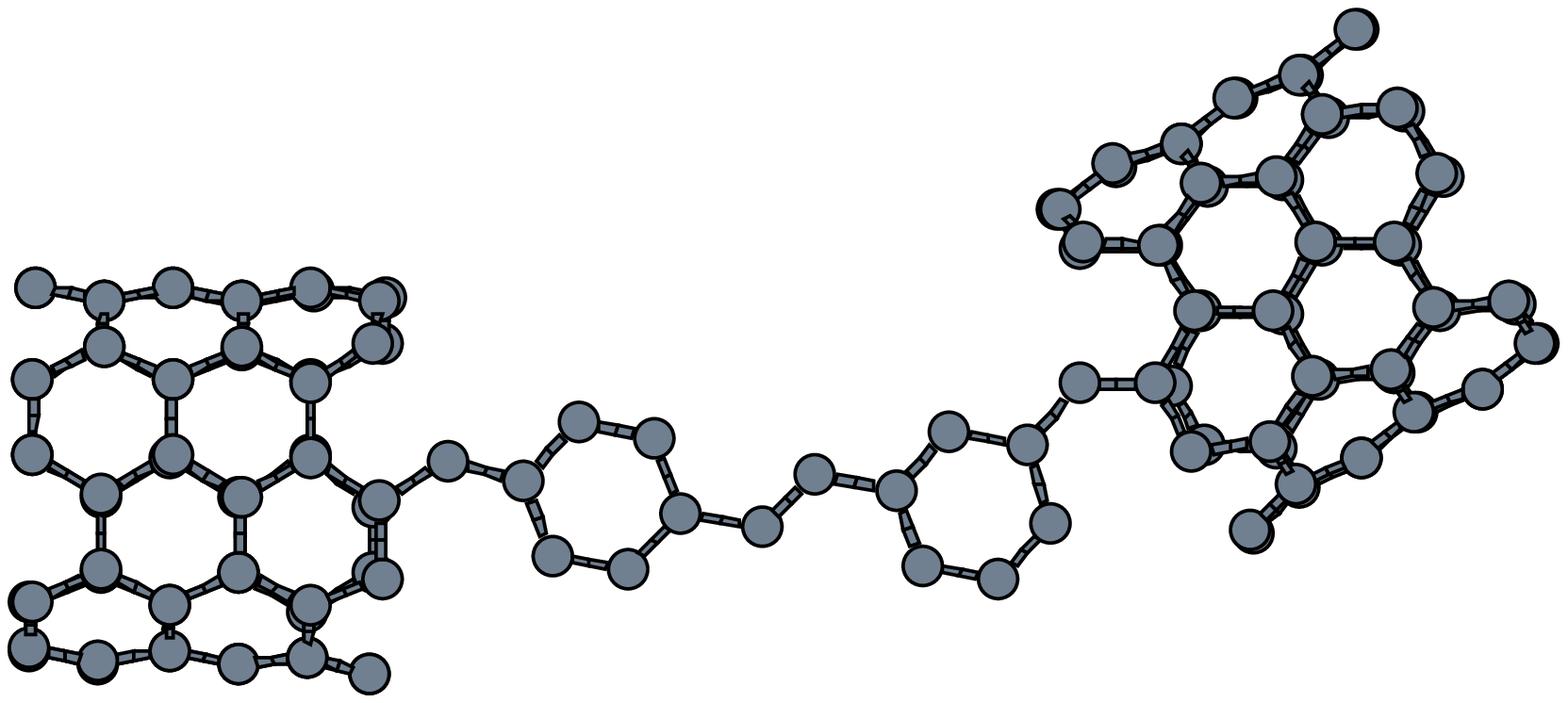, width=4.0cm, height=3.0cm}}
}
\end{minipage}
\hfill
\begin{minipage} [l] {2.9cm}
\hspace{-4.0cm}{
\raisebox {+3.95cm}[+5.75cm]
{\epsfig{figure=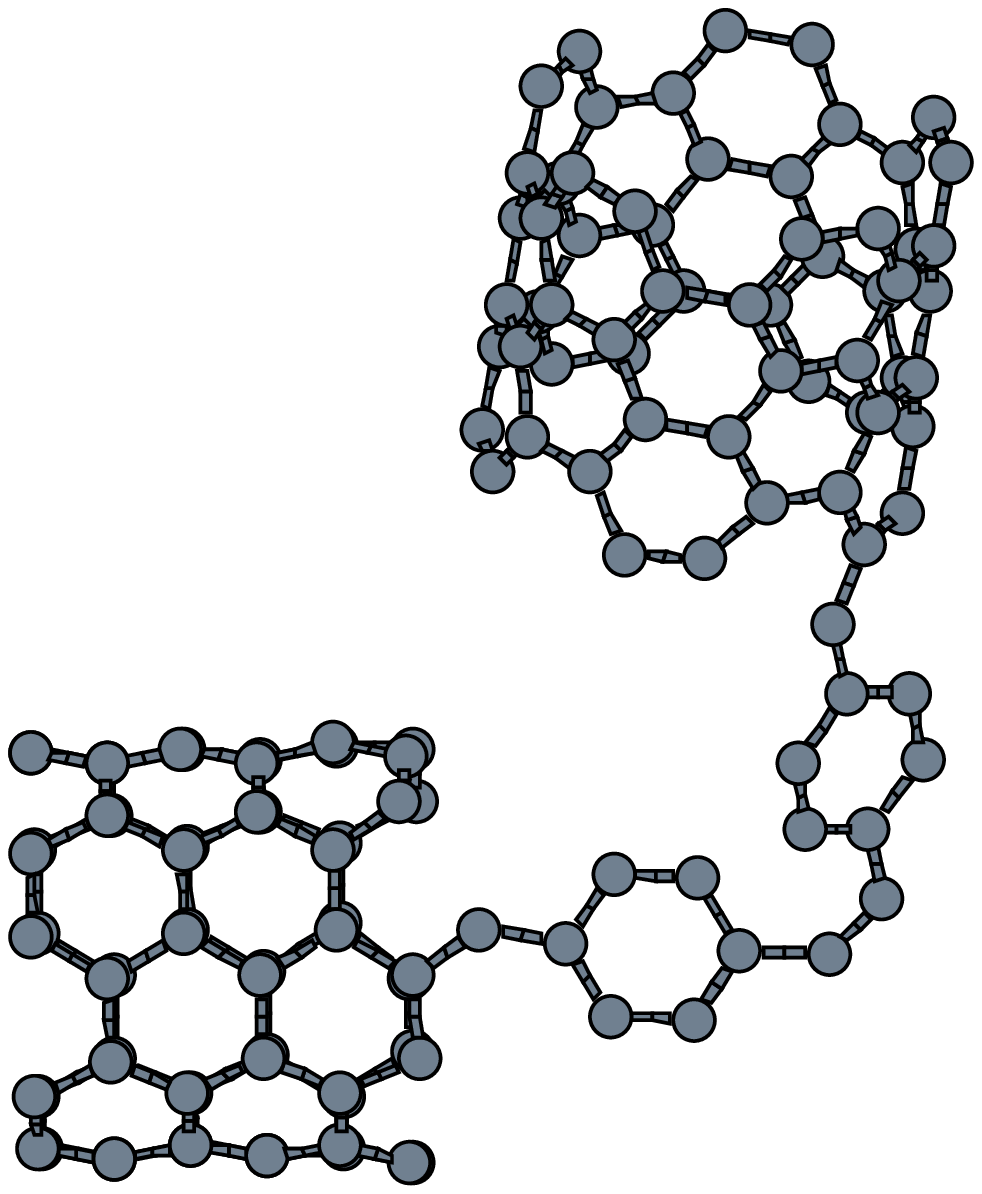, width=4.0cm, height=3.0cm}}
}
\end{minipage}

{
\vspace{-5.0cm} {
\begin{figure} [h]
\begin{center}
\epsfig{figure=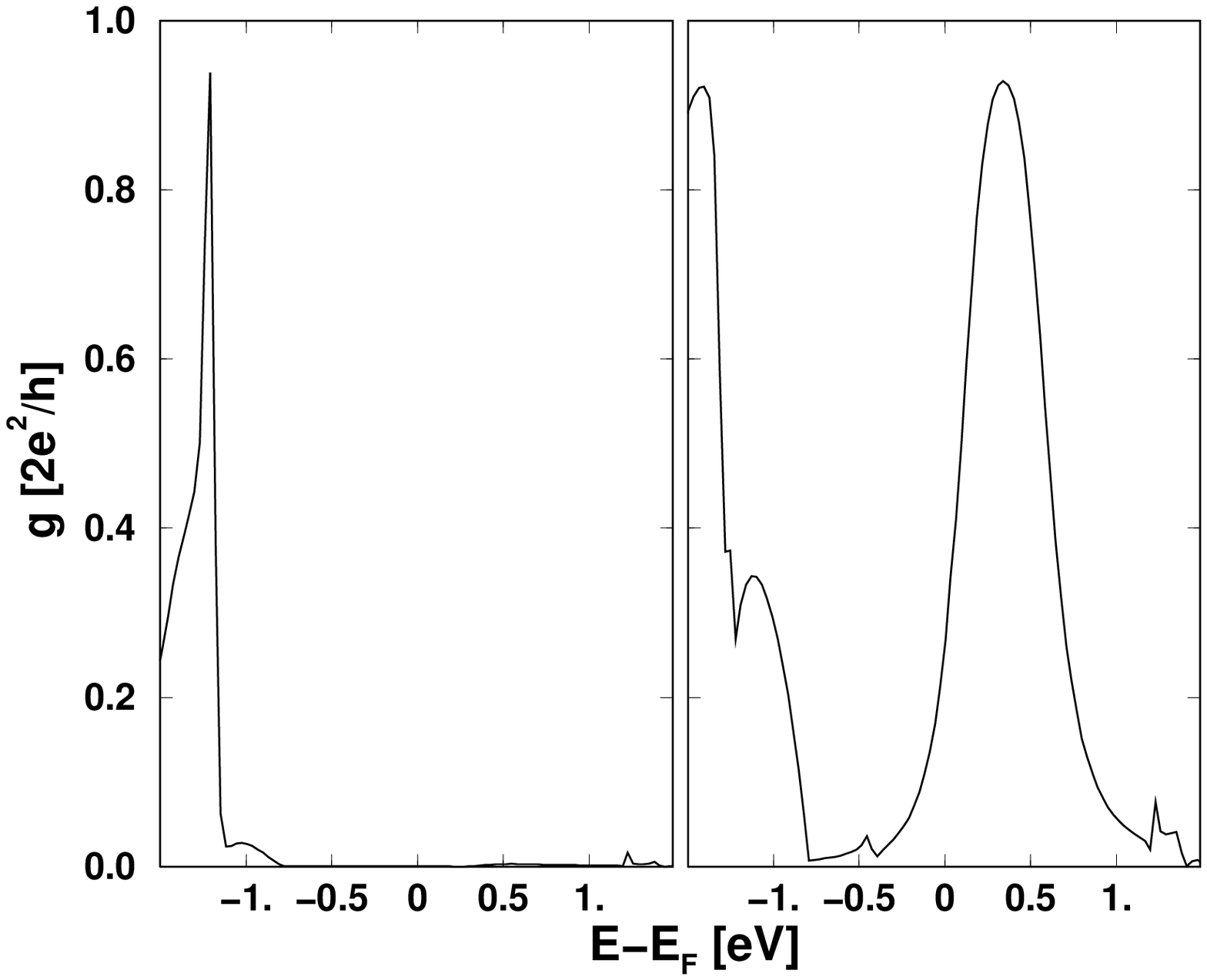, width=7.5cm, height=5.5cm}
\caption{ Conductance of {\em trans} and {\em cis} stilbene when 
changing the contact point of the molecule to the right electrode. 
}
\end{center}
\end{figure}
}}

{\it Summary}. -  Electronic transport 
through stilbene isomers  within a carbon based 
device has been investigated. A perfect interference blockade has been 
found. 
Also, we have shown that similarities in the electronic structure of both isomers 
do not necessarily imply similar 
conductance spectra, the interfacial effects being decisive 
in controlling electron transport. 
As a result, both isomers can be distinguished via transport experiments only  
if the contact geometries are appropriately chosen. 
Moreover, apart from conductance switching via photoisomerization 
the blockade effect suggests 
an alternative switching mechanism by a controlled manipulation of the 
molecule-electrode interface. 

\acknowledgments
This study was supported by the Deutsche Forschungs\-gemeinschaft
through the Forschergruppe ``Nanostrukturierte Funktionselemente
in makroskopischen Systemen''. RG thanks G. Fagas for useful comments.


\begin{thebibliography}{99}
\bibitem{comment0}
We have also tested methylene radicals, but upon relaxation
one of the H-atoms in the CH$_2$ group is removed and binds to a CNT surface atom.
\end{thebibliography}
\end{document}